\documentstyle[floats,twocolumn,epsfig,aps]{revtex}
\def\>{\rangle}
\def\<{\langle}
\def\I{\mid}

\def\ket#1{\I #1 \>}
\def\ket!#1{\I\! #1 \>}
\def\braket!#1#2{\< #1 \I #2\>}

\def\beq{\begin{equation}}
\def\eeq{\end{equation}}
\def\bea{\begin{eqnarray}}
\def\eea{\end{eqnarray}}
\def\nncr{\nonumber\\}

\def\eg{{\it e.g.}~}

\begin{document}

\twocolumn[\hsize\textwidth\columnwidth\hsize\csname @twocolumnfalse\endcsname

\title{NON-SEQUENTIAL BEHAVIOR OF THE WAVE FUNCTION}

\author{Shahar Dolev\thanks{email: shahar\_dolev@email.com}, Avshalom C. 
Elitzur\thanks{email: avshalom.elitzur@weizmann.ac.il}}
\address{Unit for Interdisciplinary Studies, Bar-Ilan University \\ 52900 Ramat-
Gan, Israel.}

\maketitle

\begin{abstract}
\noindent
An experiment is presented in which the alleged progression of a photon's wave 
function is ``measured'' by a row of superposed atoms. The photon's wave 
function affects only one out of the atoms, regardless of its position within 
the row, thereby manifesting not only non-local but also non-sequential 
characteristics. It also turns out that, out of $n$ atoms, each one has a 
probability which is higher than the normal $1/n$ to be the single affected one. 

\end{abstract}


\vspace{0.5cm}

]


\footnote[0]{$^*$email: shahar\_dolev@email.com}
\footnote[0]{$^\dagger$email: avshalom.elitzur@weizmann.ac.il }


\section {Introduction}

\begin{figure}
\begin{center}
\includegraphics[scale=0.5]{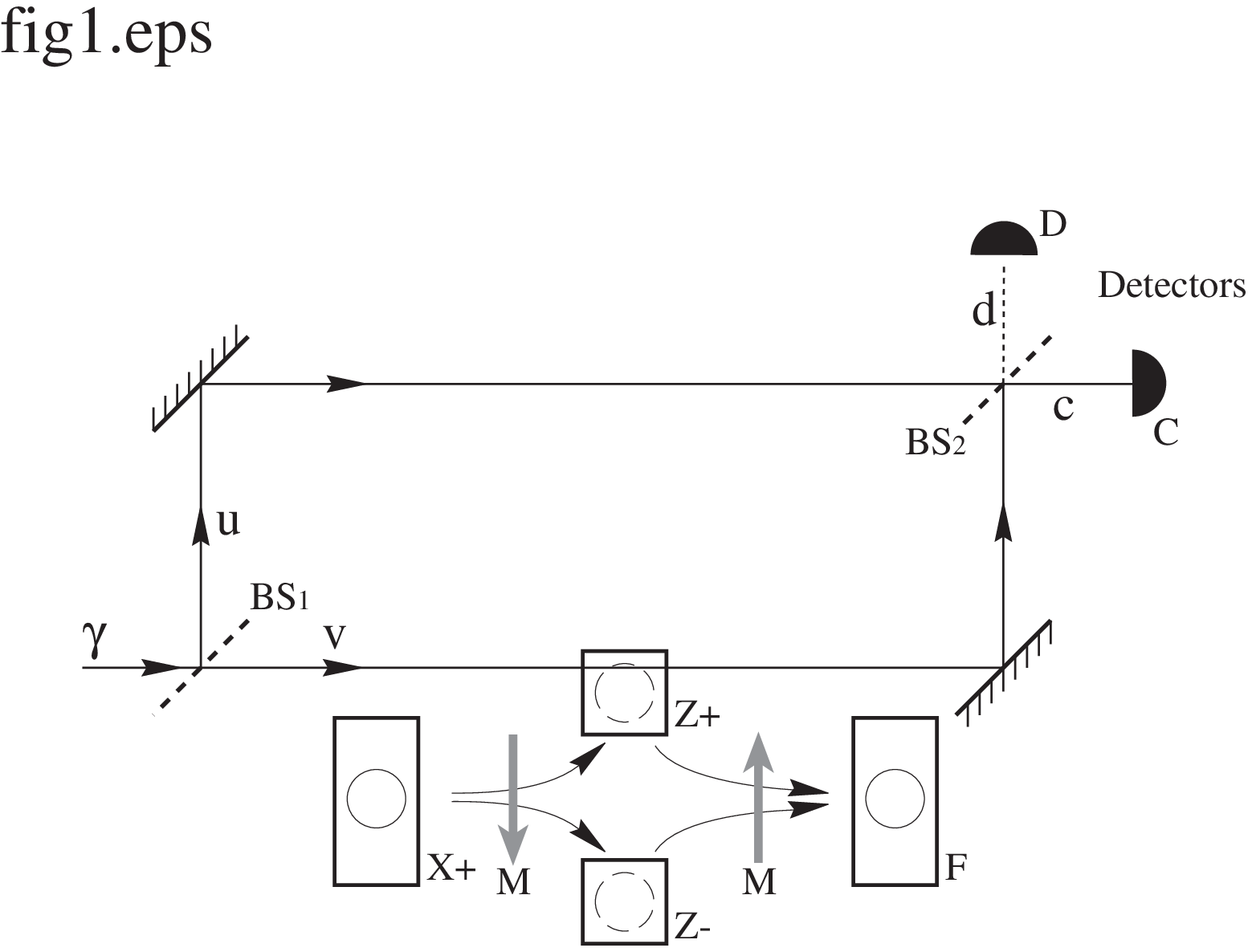}
\label{fig:1}
\vspace{0.3cm}
\caption{A photon in a Mach-Zehnder Interferometer interacting with a superposed 
atom.}
\end{center}
\end{figure}

When a single photon goes through a Mach-Zehnder Interferometer, its behavior 
indicates that it has somehow traversed both arms. However, when its position is 
measured during this passage, it turns out to have traversed only one arm. This 
is one of the notable manifestations of the measurement problem, for which 
several competing interpretations have been proposed. These can be crudely 
divided into two groups: ``collapse'' (\eg Copenhagen, GRW) and ``non-collapse'' 
(\eg Guide Wave, Many Worlds) interpretations.

Both groups, however, share one assumption. The photon -- whether in the form of 
wave-plus-particle or of a wave function evenly spread over all available 
positions - is believed to proceed from the source to the detector sequentially 
through space-time. Hence, if a few objects are placed along its path, the 
photon is expected to interact with them one after another, according to the 
order of their positions. 

In this Letter we show an experiment in which space-time sequentiality does not 
hold.

\section {Mutual IFM}

Interaction-Free Measurement (IFM) \cite{Avs1} highlights the way two 
interferometer arms, or even a myriad of them \cite{part}, are ``felt'' by a 
single particle. Its essence lies in an exchange of roles: the quantum object, 
rather than being the subject of measurement, becomes the measuring apparatus 
itself, whereas the macroscopic detector (or super-sensitive bomb in the 
original version) is the object to be measured. 
In their paper  \cite{Avs1}, Elitzur and Vaidman (EV) mentioned the possibility 
of an IFM in which both objects, the measuring and the measured, are single 
particles, in which case even more intriguing effects can appear. This 
proposition was taken up in a seminal paper by Hardy \cite{Hardy1,Hardy2}. He 
considered an EV device (Fig.\ 1) where a single photon traverses a Mach-Zehnder 
Interferometer (MZI) and interacts with an atom in the following way: A spin 
$^1\!/_2$ atom is prepared in a spin state $\ket!{X+}$ and split by a non-
uniform magnetic field $M$ into its Z components. The box is then carefully 
split into two, holding the $\ket!{Z+}$ and $\ket!{Z-}$ parts while preserving 
their superposition state: 
\beq
\Psi=\ket!{\gamma} \cdot \frac{1}{\sqrt{2}}(i\ket!{Z+}+\ket!{Z-}).
\eeq
Now let the photon be transmitted by $BS_1$:
\beq
\Psi=\frac{1}{2} \cdot (i\ket!{u}+\ket!{v}) \cdot (i\ket!{Z+}+\ket!{Z-}).
\label{eq:hardy_uv}
\eeq

The boxes are transparent for the photon but opaque for the atom. The atom's 
$Z+$ box is positioned across the photon's $v$ path in such a way that the 
photon can pass through the box and interact with the atom inside in 100\% 
efficiency. Discarding all cases of the photon's absorption by the atom (25\%) 
removes the term \hbox{$\ket!{v}\ket!{Z+}$}, leaving:
\beq
\Psi=\frac{1}{2} \cdot (-\ket!{u}\ket!{Z+} + i\ket!{u}\ket!{Z-} + 
\ket!{v}\ket!{Z-}).
\eeq
Next, let us reunite the photon by $BS_2$:
\bea
&&\ket!{v}\stackrel{BS_2}{\longrightarrow}\frac{1}{\sqrt{2}}\cdot 
(\ket!{d}+i\ket!{c})\\
&&\ket!{u}\stackrel{BS_2}{\longrightarrow}\frac{1}{\sqrt{2}}\cdot 
(\ket!{c}+i\ket!{d}),
\eea
so that
\beq
\Psi=\frac{i}{\sqrt{2}^3}\cdot [\ket!{c}\cdot (i\ket!{Z+}+2\ket!{Z-})-
\ket!{d}\ket!{Z+}].
\eeq

Once the photon reaches one of the detectors, the atom's $Z$ boxes are joined 
and a reverse magnetic field $-M$ is applied to bring it to its final state 
$\ket!{F}$. Measuring $F$'s X spin gives:
\bea
\Psi&=&\frac{1}{4}\cdot \ket!{d}\cdot (-i\ket!{X+}+\ket!{X-})\nncr
&&\quad+\frac{1}{4}\cdot \ket!{c}\cdot (-3\ket!{X+}+i\ket!{X-}).
\label{eq:hardy_cd}
\eea

Here, it can happen that the photon hits detector $D$, while the atom is found 
in a final spin state of $\ket!{X-}$ rather than its initial state $\ket!{X+}$. 
In such a case, both particles performed IFM on one another, destroying each 
other's interference. Nevertheless, the photon has not been absorbed by the 
atom, so no interaction between the photon and the atom seems to have taken 
place. 

Hardy's analysis revealed the striking consequence of this result: The atom can 
be regarded as EV's ``bomb'' as long as it is in a superposition, whereas a 
measurement that forces it to assume a definite Z spin (to ``collapse'') amounts 
to ``detonating'' it. However, the photon's hitting detector $D$ indicates that 
it has been disturbed too. And yet, in the absence of absorption, no interaction 
seems to have occurred between it and the atom. That means that the photon has 
traversed the $u$ arm of the MZI while ``detonating'' the atom on the other arm, 
forcing it to assume (as measurement indeed confirms) a definite $Z+$ spin!

Hardy argued that this case supports the guide-wave interpretation of QM. His 
reasoning was that the photon took the $u$ arm of the MZI while its accompanying 
empty wave took the $v$ arm and broke the atom's superposition. However, Clifton 
\cite{Cli} and Pagonis \cite{Pag} argued that the result is no less consistent 
with the ``collapse'' interpretation. Griffiths \cite{Gri}, employing the 
``consistent histories'' interpretation, argued that the result indicates that 
the particle might have taken the $v$ arm as well, and Dewdney {\it et.~al.~} 
\cite{Dew} reached the same conclusion using Bohmian mechanics.

To show the inadequacy of all the above analyses, let us reconsider Hardy's 
experiment with a slight yet crucial addition. Let a macroscopic object be 
placed after the atom on the $v$ arm of the photon MZI (``B'' on Fig.\ 1). Here 
Eq.\ (\ref{eq:hardy_uv}) becomes:
\beq
\Psi=\frac{i}{2} \cdot \ket!{u} \cdot (i\ket!{Z+}+\ket!{Z-}),
\eeq
and consequently Eq.\ (\ref{eq:hardy_cd}) changes into
\beq
\Psi=-\frac{1}{2} \cdot (\ket!{d}+i\ket!{c}) \cdot \ket!{X+}.
\eeq

The atom has retained its $X+$ state, indicating that the peculiar effect Hardy 
pointed out will appear only if the two halves of the wave function are allowed 
to reunite. In other words, the alleged guide wave or collapse will not exert 
their effect unless path $v$ is allowed, {\it later}, to reach $BS_2$. This 
defiance of ordinary temporal notions will become more prominent in what 
follows. 

We will now point out a more peculiar effect of the wave function for which all 
the above interpretations, due to their sequentiality assumption, seem to be 
insufficient. 

\section {IFM with one photon and several atoms}
\label{sec:several}

\begin{figure}
\begin{center}
\includegraphics[scale=0.5]{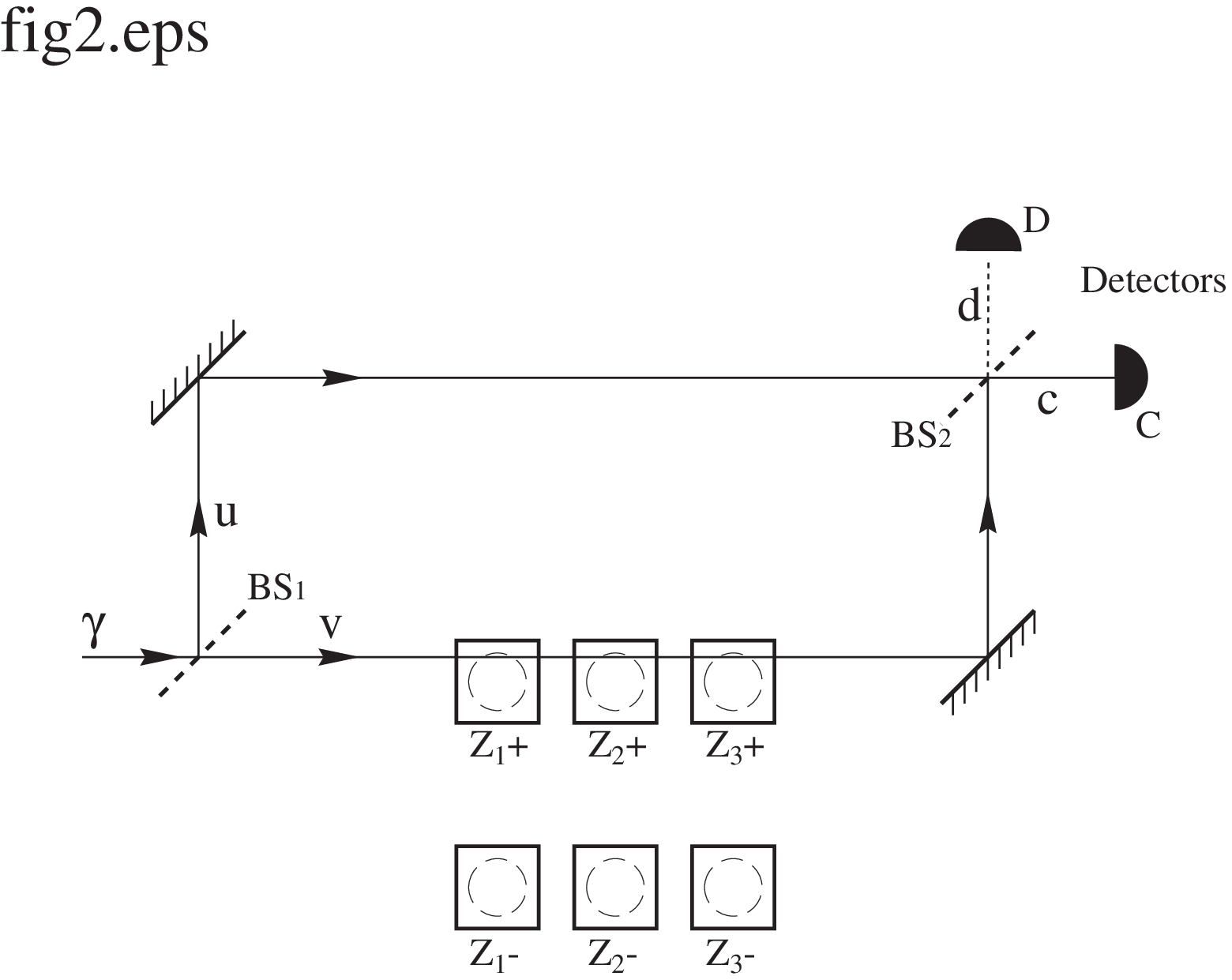}
\label{fig:2}
\vspace{0.3cm}
\caption{One photon MZI with several interacting atoms.}
\end{center}
\end{figure}

Consider the setup given in Fig.\ 2. Here too, one photon traverses the MZI, but 
now it interacts with three superposed atoms rather than one. Formally:
\beq
\Psi=\ket!{\gamma}\ket!{X_1+}\ket!{X_2+}\ket!{X_3+}.
\eeq
After the photon's passage through $BS_1$ and the atoms' splitting into their Z 
spins:
\bea
\Psi&=&\frac{1}{4}\cdot (i\ket!{u}+\ket!{v})\cdot (i\ket!{Z_1+}+\ket!{Z_1-
})\nncr
&&\quad\cdot(i\ket!{Z_2+}+\ket!{Z_2-})\cdot (i\ket!{Z_3+}+\ket!{Z_3-}).
\eea
As in the previous experiment, we discard all the cases (44\%) in which 
absorption occurred:
\bea
\Psi&=&-\frac{1}{4}\cdot [-\ket!{v} \ket!{Z_1-} \ket!{Z_2-} \ket!{Z_3-} \\
&& +\ket!{u} (  
         +i \ket!{Z_1+} \ket!{Z_2+} \ket!{Z_3-} 
         +i \ket!{Z_1+} \ket!{Z_2-} \ket!{Z_3+} \nncr
&& \quad +  \ket!{Z_1+} \ket!{Z_2-} \ket!{Z_3-} 
         +i \ket!{Z_1-} \ket!{Z_2+} \ket!{Z_3+} \nncr
&& \quad +  \ket!{Z_1-} \ket!{Z_2+} \ket!{Z_3-} 
         +  \ket!{Z_1-} \ket!{Z_2-} \ket!{Z_3+} \nncr
&& \quad -i \ket!{Z_1-} \ket!{Z_2-} \ket!{Z_3-} 
         -  \ket!{Z_1+} \ket!{Z_2+} \ket!{Z_3+}
    ) ]. \nonumber
\eea
Now let us pass the photon through $BS_2$ and select only these cases in which 
it has lost its interference, hitting detector $D$:
\bea
\Psi &=& \frac{1}{4\sqrt{2}} \cdot \ket!{d} \nncr
&& \cdot (i \ket!{Z_1+} \ket!{Z_2+} \ket!{Z_3+}
          + \ket!{Z_1+} \ket!{Z_2+} \ket!{Z_3-} \nncr
&& \quad  + \ket!{Z_1+} \ket!{Z_2-} \ket!{Z_3+}
         -i \ket!{Z_1+} \ket!{Z_2-} \ket!{Z_3-} \nncr
&& \quad  + \ket!{Z_1-} \ket!{Z_2+} \ket!{Z_3+}
         -i \ket!{Z_1-} \ket!{Z_2+} \ket!{Z_3-} \nncr
&& \quad -i \ket!{Z_1-} \ket!{Z_2-} \ket!{Z_3+} ).
\label{eq:atomsatz}
\eea
Measuring the 3 atoms' spins now will yield, with a uniform probability, all 
possible results, except the case where all the atoms are found in their 
$\ket!{Z-}$ boxes, which will never occur. 

Reuniting the atoms' $Z$ boxes and measuring their X spin will yield all 
possible combinations of $X+$ and $X-$ in uniform probability, except the case 
of all three atoms measuring $X+$ which has a higher probability. This is not 
surprising, as these atoms are supposed to have interacted either with the guide 
wave, or with the real particle itself (see \cite{Hardy1}) or with the 
uncollapsed wave function \cite{Cli,Pag}. 

Let us, however, return to the stage before uniting the $Z$ boxes (as per Eq.\ 
(\ref{eq:atomsatz})). We know that at least one atom must be in the $\ket!{Z+}$ 
box to account for the loss of the photon's interference. Let us, then, measure 
atom 2's spin, and proceed only if it is found to be $\ket!{Z_2+}$ (56\% of the 
cases):
\bea
\Psi=\frac{1}{4\sqrt{2}}&& \cdot \ket!{d} \cdot (-i \ket!{Z_1-} + \ket!{Z_1+} ) 
\nncr
&& \cdot \ket!{Z_2+} \cdot (i \ket!{Z_3+} + \ket!{Z_3-} ).
\label{eq:a}
\eea
Now unite the Z boxes of atoms 1 and 3 and apply the reverse magnetic field $-
M$:
\beq
\Psi=\frac{1}{2\sqrt{2}} \cdot \ket!{d} \cdot \ket!{X_1+} \cdot \ket!{Z_2+} 
\cdot \ket!{X_3+}.
\label{eq:b}
\eeq

Surprisingly, these atoms will {\it always} exhibit their original spin 
undisturbed, just as if no photon has ever interacted with them. In other words, 
only one atom is affected by the photon in the way pointed out by Hardy, but 
that atom does not have to be the first one, nor the last; it can be any one out 
of the atoms. The other atoms, whose half wave functions intersected the MZI arm 
before or after that particular atom, remain unaffected.

We can prove, however, that although atom 3 seems to be totally unaffected by 
the photon, {\it something} must have passed through it. As in the previous 
section, let a macroscopic object be placed further along the $v$ route, after 
the three atoms (object ``B'' on Fig.\ 2). The above results will never show up. 
Here, all the atoms will give either $Z-$ (when the photon hits the obstacle), 
or $X+$ (when it does not). Hence, something must have passed through all three 
atoms, yet it has left the first and last unaffected.

The next result will deal the final blow on any realistic account in which a 
particular atom is affected by the photon at the moment of their interaction. We 
noted above that if we pick one atom, measure its position and find it to reside 
in the $Z+$ box, then that measurement will disentangle the two other atoms and 
their spins will reveal no trace of interaction with the photon. One might think 
that there is, prior to measurements of the atoms, one particular atom that 
``has been'' affected, and that the experimenter only has to be lucky to pick up 
that ``right'' atom that yields $\ket!{Z+}$. Not so: rather than the normal 33\% 
probability to find the ``right'' atom, expected when there are 3 atoms, the 
probability is 56\%! 

In other words, it is the very choice of an atom by the experimenter, regardless 
of its position within the row, that increases the probability for that atom to 
be ``the only atom that has been affected by the photon.'' And once this atom 
gives this result, the other atoms will become disentangled.\footnote{ No 
superluminal communication is entailed. In all cases in which the particular 
atom is not detected in the intersecting path, the probability for one of the 
other atoms to reside in that path increases to 1. The overall result is 
Lorentz-invariant. Still, the correlation is Bell-like.} 

Note that the above analysis does not depend on the number of atoms or the index 
of the tested atom. For $n$ atoms, the probability for any atom to be ``the 
right atom'' is $P={2^{n-1}+1 \over 2^n}$ instead of the expected $P=1/n$, 
reaching $1/2$ as $n$ increases.

Finally, we can demonstrate that, although all the atoms but the middle one 
reveal no indication that they have ever interacted with the photon, something 
physical must have passed through them at the right time. Let us place the atoms 
within sealed boxes, with apertures at the $v$ path, which open only for the 
minute interval during which the photon's wave function is supposed to pass 
through them. The slightest failure in the timing of any aperture's opening will 
ruin the predicted result.

\section{Conclusions}

It is not surprising that several atoms, which have interacted with one photon, 
yield strictly correlated results (one entangled, the others disentangled). But 
it is the attempt to reconstruct a comprehensible evolution from these 
correlations that gives a highly counterintuitive picture: A single photon's 
wave function seems to ``skip'' a few atoms that it encounters, then disturb the 
$m^{th}$ atom, and then again leave all next atoms undisturbed. Ordinary 
concepts of motion, which remain implicit in both ``guide wave'' and 
``collapse'' interpretations, are inadequate to explain this behavior. Even more 
exotic interpretations of QM, such as those invoking advanced actions 
\cite{Whe,Cra,Rie,Bea} or tunneling, can be shown to be inadequate for 
explaining this result.

The most prudent description of this result is that a wave function, when 
interacting with a row of other wave functions one after another, does not 
comply with ordinary notion of causality, space and time.

\acknowledgments{We thank Yakir Aharonov, David Tannor and Terry Rudolph for 
very helpful comments. It is a pleasure to thank Anton Zeilinger and all 
participants of the Quantum Measurement Conference at the Schr\"odinger 
Institute in Vienna for enlightening discussions.}

\bibliographystyle{unsrt}

\end{document}